\documentclass[12pt, epsfig]{article}
\usepackage{epsfig}
\usepackage{amssymb}
\usepackage{amsfonts}
\usepackage{bbm}
\usepackage{amsmath}
\usepackage[latin1]{inputenc}
\usepackage{layout}

\newskip\humongous \humongous=0pt plus 1000pt minus 1000pt

\newif\ifdtup

\catcode`\@=11

\@addtoreset{equation}{section}
\def\theequation{\thesection.\arabic{equation}}

\def\@normalsize{\@setsize\normalsize{15pt}\xiipt\@xiipt
\abovedisplayskip 14pt plus3pt minus3pt%
\belowdisplayskip \abovedisplayskip
\abovedisplayshortskip \z@ plus3pt%
\belowdisplayshortskip 7pt plus3.5pt minus0pt}

\def\small{\@setsize\small{13.6pt}\xipt\@xipt
\abovedisplayskip 13pt plus3pt minus3pt%
\belowdisplayskip \abovedisplayskip
\abovedisplayshortskip \z@ plus3pt%
\belowdisplayshortskip 7pt plus3.5pt minus0pt
\def\@listi{\parsep 4.5pt plus 2pt minus 1pt
     \itemsep \parsep
     \topsep 9pt plus 3pt minus 3pt}}

\relax

\catcode`@=12

\catcode`\@=11

\def\section{\@startsection{section}{1}{\z@}{3.5ex plus 1ex minus
   .2ex}{2.3ex plus .2ex}{\large\bf}}

\def\thesection{\arabic{section}}
\def\thesubsection{\arabic{section}.\arabic{subsection}}

\def\appendix{\setcounter{section}{0}
 \def\thesection{Appendix \Alph{section}}
 \def\thesubsection{\Alph{section}.\arabic{subsection}}
 \def\theequation{\Alph{section}.\arabic{equation}}}

\def\YGrule{0.4}   
\def\YGbox{6.5}    
\def\SymBoxes#1#2#3#4{\newdimen\un@t \un@t#3%
\raisebox{#1}{\rule{#2\un@t}{#4}\hskip-#2\un@t
\@tempdimb\un@t \advance\@tempdimb by-#4\@tempcntb#2\relax%
\@whilenum{\@tempcntb>0}\do{
\rule{#4}{\un@t}\hskip\@tempdimb \advance\@tempcntb by\m@ne}%
\hskip-#2\un@t \rule[\un@t]{#2\un@t}{#4}%
\rule[\un@t]{#4}{#4}\hskip-#4
\rule{#4}{\un@t}}\hskip-#4}                
\def\Young{\@ifnextchar[{\@Young}{\@Young[0]}}
\def\@Young[#1]#2{\newdimen\YG@unit \YG@unit\YGbox pt%
\newdimen\h@ight \h@ight#1\YG@unit \@tempcnta-1\relax
\@tfor\c@ount:=#2\do{\advance\@tempcnta by\@ne}
\@tempdima\@tempcnta\YG@unit%
\advance\h@ight by\@tempdima\relax     
\@tfor\c@ount:=#2\do{\SymBoxes{\h@ight}{\c@ount}{\YG@unit}{\YGrule pt}%
\@tempdima-\c@ount\YG@unit \hskip\@tempdima%
\advance \h@ight by -\YG@unit}         
\@tempdima\YG@unit \multiply\@tempdima by\@car#2\@nil %
\hskip\@tempdima}                      
\def\YoungTab{\@ifnextchar[{\@YoungIdx}{\@YoungIdx[0]}}
\def\@YoungIdx[#1]{\@ifnextchar[{\@iYoungIdx[#1]}{\@iYoungIdx[#1][\@empty]}}
\def\@iYoungIdx[#1][#2]#3{%
\newdimen\YG@unit \YG@unit\YGbox pt\newdimen\YG@rule \YG@rule \YGrule pt
\newcount\c@ount \c@ount\z@ \newdimen\skip@wd \unitlength\@ne pt
\newdimen\h@ight \h@ight#1\YG@unit \@tempcnta\m@ne\relax
\@tfor\d@um:=#3\do{\advance\@tempcnta by\@ne}
\@tempdima\@tempcnta\YG@unit%
\advance\h@ight by\@tempdima\relax
\@tfor\@idxlist:=#3\do{
\@tempcnta\z@\hskip.5\YG@rule\relax
\@for\@idx:=\@idxlist\do{
\raisebox{\h@ight}{\makebox(\YGbox,\YGbox){#2$\@idx$}}
\advance\@tempcnta by\@ne}\hskip-.5\YG@rule%
\@tempdima-\@tempcnta\YG@unit \hskip\@tempdima%
\ifnum\c@ount=\z@ \skip@wd-\@tempdima\fi \relax
\SymBoxes{\h@ight}{\@tempcnta}{\YG@unit}{\YG@rule}%
\hskip\@tempdima \advance\h@ight by -\YG@unit
\advance\c@ount by\@ne}
\hskip\skip@wd}                      

\begin{document}

\newcommand{\be}{\begin{equation}}
\newcommand{\ee}{\end{equation}}
\newcommand{\bqa}{\begin{eqnarray}}
\newcommand{\eea}{\end{eqnarray}}
\newcommand{\beas}{\begin{eqnarray*}}
\newcommand{\eeas}{\end{eqnarray*}}
\newcommand{\defi}{\stackrel{\rm def}{=}}
\newcommand{\non}{\nonumber}
\newcommand{\bquo}{\begin{quote}}
\newcommand{\enqu}{\end{quote}}
\def\de{\partial}
\def\Tr{ \hbox{\rm Tr}}
\def\tr{ \hbox{\rm tr}}
\def\const{\hbox {\rm const.}}
\def\o{\over}
\def\im{\hbox{\rm Im}}
\def\re{\hbox{\rm Re}}
\def\ket#1{|{#1}\rangle}
\def\bra#1{\langle {#1} |}
\def\ckt{\rangle}
\def\brc{\langle}

\def\Arg{\hbox {\rm Arg}}
\def\Re{\hbox {\rm Re}}
\def\Im{\hbox {\rm Im}}
\def\diag{\hbox{\rm diag}}
\def\longvert{{\rule[-2mm]{0.1mm}{7mm}}\,}

\begin{titlepage}
{\hfill     IFUP-TH/2011-04}
\bigskip

\begin{center}
{\large  {\bf
Nonabelian Faddeev-Niemi Decomposition of the SU$(3)$ Yang-Mills Theory
 } }
\end{center}

\bigskip
\begin{center}
{\large  J. Evslin$^{1}$, S. Giacomelli$^{3,4}$,  K. Konishi$^{2,4}$,  A. Michelini$^{3,4}$   \vskip 0.10cm
 }
\end{center}

\begin{center}
{\it   \footnotesize
Institute of High Energy Physics\\
Chinese Academy of Sciences, P.O. Box 918-4, Beijing 100039, P. R. China $^{(1)}$\\
Dipartimento di Fisica ``E. Fermi" -- Universit\`a di Pisa $^{(2)}$, \\
   Largo Bruno Pontecorvo, 3, Ed. C, 56127 Pisa,  Italy \\
Scuola Normale Superiore - Pisa $^{(3)}$,
 Piazza dei Cavalieri 7, Pisa, Italy \\
Istituto Nazionale di Fisica Nucleare -- Sezione di Pisa $^{(4)}$, \\
     Large Bruno Pontecorvo, 3, Ed. C, 56127 Pisa,  Italy
   }

\end {center}

\noindent
{\bf Abstract:}
Faddeev and Niemi (FN) have introduced an abelian gauge theory which simulates dynamical abelianization in Yang-Mills theory (YM).  It contains both YM instantons and Wu-Yang monopoles and appears to be able to describe the confining phase. Motivated by the meson degeneracy problem in dynamical abelianization models, in this note we present a generalization of the FN theory. We first generalize the Cho connection to dynamical symmetry breaking pattern SU($N+1$) $\to$ U($N$), and
subsequently try to complete the Faddeev-Niemi decomposition by keeping the missing degrees of freedom.  While it is not possible to write an on-shell complete FN  decomposition, in the case of SU($3$) theory of physical interest  we find an off-shell complete decomposition for SU($3$) $\to$ U($2$) which amounts to partial gauge fixing, generalizing naturally the result found by Faddeev and Niemi for the abelian scenario SU($N+1$) $\to$ U($1$)$^{N}$.  We discuss general topological aspects of these breakings, demonstrating for example that the FN knot solitons never exist when the unbroken gauge symmetry is nonabelian,
and recovering the usual no-go theorems for colored dyons.
\vfill

\begin{flushright}
  March 2011
\end{flushright}
\end{titlepage}

\bigskip

\hfill{}

  \section{Introduction}

In a remarkable series of papers Faddeev and Niemi \cite{FN1,FN2} have proposed a variation of SU($N+1$) Yang-Mills theory which may provide the simplest known context in which to study quark confinement. Their construction is roughly a partial gauge fixing of Yang-Mills theory \`a la 't Hooft \cite{thooft},
\be   {\rm SU}(N+1)  \to   {\rm U}(1)^{N},
\label{Abel}  \ee
which exhibits rather explicitly the magnetic degrees of freedom which are, in the standard varables $A_{\mu}^{a}$, encoded as non-local, topological properties of the configuration as a whole.  Assuming that  in the infrared the magnetic degrees of freedom become somehow dominant, Faddeev and Niemi have shown \cite{nodi1,nodi2} that these systems possess topologically stable toroidal solitons\footnote{Of course the dynamical breaking of scale invariance and generation of the mass scale $\Lambda$ is
crucial for allowing for such solitons to appear, evading Derrick's theorem.  } which could be interpreted as models of the glueballs.  Their approach might also be useful  in analyzing the interrelation between the two dynamical aspects of the system,  confinement and chiral symmetry breaking.

Recently \cite{noi,Niemi} we have shown that the Faddeev-Niemi decomposition must be handled with care, since the equations of motion deriving from their
formula are not equivalent to the original Yang-Mills equations and, if not supplemented by imposing the Gauss' constraints, give rise to solutions
which are not part of Yang-Mills theory. More precisely, the FN theory contains some solutions which are solutions of YM coupled to an external color current.  Nevertheless the overlap with Yang Mills theory is remarkable, as many known solutions
of the Yang-Mills equations of physical interest are encoded in their formula.  Also variations on Faddeev and Niemi's construction, such as those in Refs.~\cite{Shabanov,BCK,Kondo}, do not suffer from this problem, and indeed the problem does not present itself if one does not fix the constraints \cite{FNvinc}.

There is, however,  no clear evidence for the moment that the effectively abelian picture (dynamical abelianization) is valid for the real-world Quantum Chromodynamics (QCD). It is therefore important to study every venue searching for other dynamical possibilities. One of them is that the effective gauge symmetry breaking leaves some nonabelian factor unbroken. For definiteness here we focus on the symmetry breaking pattern
\be   SU(N+1)  \to   \frac{SU(N)\times U(1)}{{\mathbbm Z}_{N}} \sim U(N)\;.
\label{NonA}\ee
Note that such a nonabelian scenario is not possible in the SU(2) gauge theory, but {\it is}  possible in the case of SU($3$) Yang-Mills theory, the theory of interest for QCD.
Of course such a decomposition will necessarily bring us back to the well-known difficulties \cite{CDyons,DFHK} associated with the  nonabelian monopoles
found in the study of the gauge breaking  (\ref{NonA}) in a Higgs-type theory (i.e.,  with scalar fields in the adjoint representation),
and we do not expect our analysis here alone to be sufficient for solving them.  Examples found in various supersymmetric theories \cite{susy} suggest that understanding of nonabelian monopoles ultimately requires subtle considerations involving the renormalization group, phases of the system, the role of massless matter fields in nonabelian electromagnetic duality, and a significant interplay of magnetic monopoles and vortices \cite{noimono,Duality,Confind}.

This is not however the main interest of the present investigation.  We will restrict our attention to classical field theories,
extending Faddeev and Niemi's decomposition to the case of SU($3$) pure Yang-Mills dynamically broken to U(2), which we will simulate with a U(2) gauge theory.   Such a connection, like the FN theories, captures many of the interesting YM solutions although more generally also contains solutions which are coupled to an external current.  We feel that this case is particularly relevant for understanding confinement in real world QCD, because the the abelian breaking SU$(3)\rightarrow$U$(1)\times$U$(1)$ yields an unphysical degeneracy in the meson spectrum \cite{DouglasShenker,HananyZaffaroni}, while this breaking does not \cite{ABEKY}.   Our construction is a natural generalization of the off-shell decomposition by  Faddeev and Niemi,  corresponding to partial gauge fixing:
the unbroken U(2) gauge degrees of freedom are kept off-shell.

 In Section \ref{topsez} we discuss general topological aspects of these symmetry breaking patterns, and show how monopoles and instantons are described in this formalism.  We discuss the topological obstruction to colored dyons in this context, and show that FN knot solutions \cite{nodi1,nodi2} do not exist when the unbroken gauge symmetry is nonabelian.

 In ~\ref{transsez} we analyze in detail the transformation properties of our generalized Cho under the restricted gauge rotations. In ~\ref{genecho} we study the Cho connection for more general symmetry breaking patterns. Finally, an example of the conjugation matrix without topological obstruction is constructed  in \ref{Noobst}.

\section{SU($N$) vielbeins in the SU($N+1$) Yang-Mills theory} \label{Chosez}

\subsection{U($N$) embedding in SU($N+1$) }

Let us consider the pure SU($N+1$) Yang-Mills theory.
Define the traceless part of a rank-one   matrix
\be  M =   z\, {\bar z}  - \frac{1}{N+1} {\mathbbm 1},   \ee
where
\be   z = \left(\begin{array}{c}z^1 \\z^2 \\\vdots \\z^{N+1}\end{array}\right) \ee
is an $(N+1)$-component complex vector of unit length,
\be   {\bar z}  z =  \sum_{a=1}^{N+1}\,  {\bar z}_{a}  z^{a} =1\;.
\ee
$M$ satisfies
\be   \Tr  \,  M^{2}  =\frac{N}{N+1}, \qquad  \Tr M =0\;.
\ee
The  traceless Hermitian matrix $M$ is an element of the Lie algebra of $\mathfrak{su}(N+1)$.

Let the vector $z$ depend on the spacetime point $x$, $z=z(x)$.  $M(x)$ thus defines a space-time dependent  embedding U$(N)\hookrightarrow\ $SU$(N+1)$ where U$(N)$ is the subgroup which commutes with $M(x)$.  Such embeddings correspond to points in the manifold
\be  \mathbb{CP}^{N}={\rm SU}(N+1)/{\rm U}(N).   \label{CPN}
\ee
Alternately, as  the vectors $z_{i} $ are points in
$   S^{2N+1}\;,
$
but as   $M$ is invariant under the U(1) rotation
$  z \sim   e^{i \alpha}  z$,
$M\in\mathbb{CP}^N={S^{2N+1}}/{U(1)} $.

The $(N+1)\times (N+1)$  Hermitian matrix,
$M$ has an eigenvector $z$
with eigenvalue
$ {N}/{N+1}$,
 and $N$ orthonormal eigenvectors $e_i$ orthogonal to $z$ \footnote{A somewhat analogous idea of introducing ``zweibeins''  in the planes orthogonal to the Abelian, Cartan subalgebra directions, has been used to introduce the dual gauge symmetries in \cite{FNZwei}.},
 \be     e_{i}  = \left(\begin{array}{c}e_{i}^1 \\  e_{i}^2 \\\vdots \\e_{i}^{N+1}\end{array}\right)     \qquad i=1,2,\ldots N, \qquad   {\bar z}  \cdot   e_{i} =0,  \quad {\bar e}^{i} \cdot  e_{j} =  \delta_{j}^{i},  \label{orthnorm}
 \ee
with the degenerate eigenvalue,
$ - {1}/{N+1}\;$.
The matrix  $e_{i}^{a}$, ($i=1,2,\ldots N$, $a=1,2,\ldots N+1$)  transforms as an $N+1$ vector under the left action of SU($N+1$) as well as an $N$ vector under the right action of the subgroup SU($N$) $\subset$ U($N$) .


Any Hermitian matrix $M$ can be diagonalized by  a unitary matrix,  made from its  $N+1$ orthonormal eigenvectors,
\be U=  \left(    \left(\begin{array}{c}   \\ z  \\  \\  \end{array}\right)   \left(\begin{array}{c}   \\ e_{1}  \\  \\  \end{array}\right)   \cdots   \left(\begin{array}{c}   \\ e_{N}  \\  \\  \end{array}\right)   \right),
\ee
so that
\be   U^{\dagger}   M  U =   \sqrt {\frac{2N}{N+1}} \,  T^{(0)}, \qquad   T^{(0)}= \frac{1}{\sqrt{2N(N+1)}}   \left(\begin{array}{cc}N & 0 \\0 & -   {\mathbbm 1}_N\end{array}\right),  \label{MtoU}
\ee
is the U(1)$ \subset$ SU($N+1$) generator in some fixed direction.   Accordingly, $M$ can be written as the conjugacy class of $T^{(0)}$,
\be         M =  \sqrt {\frac{2N}{N+1}} \,   U  \, T^{(0)}\,    U^{\dagger}, \qquad   \Tr \,  (T^{(0)})^{2} =\frac{1}{2}\;, \label{UtoM}
\ee
showing manifestly that  $U$  (which determines $M$ uniquely)  determines an element of the $2N$-parameter  coset space   Eq.~(\ref{CPN}).
However $M$ does not determine $U$ uniquely, but only up to a choice of basis of the $e_i$.
Note that the vectors $z$ and $\{ e_{i}^{a} \}$ satisfy, in addition to the orthonormality conditions, the completeness relation
\be    \ket{z}  \bra{z} + \sum_{i}  \ket{e_{i}}  \bra{e_{i}}  =  {\mathbbm 1}, \qquad {\rm or}
\qquad  z^{a}{\bar z}_{b} +  \sum_{i}  {e_{i}^{a}}  {\bar  e}^{i}_{b} = \delta^{a}_{b}\;,
\ee
so that $M$ can be expressed also as
\be  M =   \frac{N}{N+1} {\mathbbm 1}  -   \sum_{i}    e_{i}\, {\bar  e}^{i}  =   \frac{N}{N+1} z \bar z  -   \frac{1}{N+1}  \sum_{i}    e_{i}\, {\bar  e}^{i} \;,   \label{clearly}
\ee
and so on.

%

%
\subsection {The Cho connection  \label{sec:Cho}}

The generalization of the Cho connection suitable for  the symmetry breaking pattern (\ref{CPN})  can be written down by using the
SU($N$) vielbeins introduced above.
 Following the original idea of Cho~\cite{DuanGe,Cho1,Cho2},  we seek for a reduced Yang Mills connection satisfying the condition
 \be      D_{\mu}  \, M =   \de_{\mu}M - i[ A_{\mu} , M]  =   0\;, \label{ChoCond}
\ee
meaning that there is a covariantly constant $M$  signaling an effective, dynamically generated  Higgs mechanism \footnote{Recently a nonabelian Cho connection similar to ours has been used in \cite{Kondonew} to study the Wilson loop  of the ${\rm SU}(3)$  Yang-Mills theory.
}.
The solution of this condition turns out to be of the form
\be    A_{\mu}  = C_{\mu}^{(0)}  \, M  +  i \,  [M, \de_{\mu} M] +  { B}_{\mu}  - E_{\mu}
\label{Ansatz}\ee
where
\be   { B}_{\mu\, b}^{a} =   e_{i}^{a}  B^{i}_{\mu\, j}  {\bar e}^{j}_{b} ,
\qquad      E_{\mu\, b }^{a}   =  i \, e_{i}^{a} \,  \left[ {\bar e}^{i}_{c} \de_{\mu} e_{j}^{c} -  \delta^{i}_{j}\frac{1}{N}  \Tr  ({\bar e} \de_{\mu} e)\right]  \, {\bar e}^{j}_{b}\;.
\ee
$ C_{\mu}^{(0)} $  is the abelian gauge field and $B^{i}_{\mu\, j} $  are the components of the gauge fields of the local (dual) SU($N$) symmetry.
It can be easily checked that (\ref{Ansatz}) indeed satisfies the gauge condition (\ref{ChoCond}).

Although $B_{\mu}$ and $E_{\mu}$ can be combined into a single term  (they are both of the form $  e \,( \ldots) \, {\bar e}$)  and only $B^{i}_{\mu\, j} $ contains new degrees of freedom with respect to the vielbeins  $\{ e_{i}^{a} \}$,   with this splitting the decomposition  (\ref{Ansatz}) of $A_{\mu}$ is form-invariant with respect to U($N$), 
as will be seen below. The term $E_{\mu}$ is basically telling us that as we vary the embedding of U($N$) inside SU($N+1$), the SU($N$) subgroup rotates. This feature is due to the nonabelianess of the ``residual" group and is unavoidable.

The original $A_{\mu} $ fields contain on-shell
\be    2 \, ((N+1)^{2}-1 ) =  2 N^{2} + 4N\;
\label{OSDG} \ee
degrees of freedom.
The new variables contain
\[     {\#} (B^{i}_{\mu\, j}) =   2 \, (N^{2}-1 ), \qquad   {\#} (e_{i}^{a}) =  2N, \qquad     {\#} (C_{\mu}) =  2,
\]
 i.e., the total of
\[   2 \, (N^{2}-1 )+ 2N+2 =  2 N^{2} + 2N\;
\]
on-shell degrees of freedom, $2N$ less than  (\ref{OSDG}).  More about this later.

The counting of the degrees of freedom in the vielbeins $e$ works as follows.
Apparently $e$ contains $2 N (N+1)$
parameters.   Taking into account the {\it invariance}  of $A_{\mu}$ under the local SU($N$) transformations  Eq.~(\ref{simul}),    the unobservable phase of $e$
and the orthonormality condition among $e$'s
\[   {\bar e}^{i} \cdot  e_{j} =  \delta_{j}^{i},
\]
the net number of degrees of freedom of $e$ is found to be
\be      2 N (N+1) -  (N^{2}-1) - 1  -  [\, N(N-1) +N\, ] = 2N. \label{counting}
\ee
$z$  (or $M$) is recovered uniquely from $e$  (see Eq.~(\ref{clearly})).  Or equivalently one can use simply $z$  ($2N$ free parameters).

\subsection{Dual  U($N$) transformations \label{UNCho}}

    We now show  how the U($N$) gauge transformations orthogonal to  $M$  act on the Cho connection, (\ref{Ansatz}).
First consider  the U($1$) transformation around the $M$ direction,
\be  U =  e^{i \alpha M}  =    e^{i \alpha  \frac{N}{N+1} }  z {\bar z}   +   e^{- i \alpha  \frac{1}{N+1} }    e {\bar e}\;
\label{u1tr}\ee
where a convenient form of $M$ (Eq.~(\ref{clearly}) has been used.)
The fields $B_{\mu}$ and $E_{\mu}$ are easily seen to be invariant.   By studying quantities  $U \, \partial_{\mu}M \, U^{\dagger}$,   $U  \,i \,[M,\partial_{\mu}M] U^{\dagger}$, and
$i\,\partial_{\mu}UU^{\dagger}$  (Appendix A)  and collecting the results,  one ends up with a simple transformation law,
\bqa  A_{\mu}  &  \rightarrow  & A^{U}_{\mu}\equiv UA_{\mu}U^{\dagger}-i\partial_{\mu}UU^{\dagger}    \nonumber  \\
&=&  C_{\mu}M+Ui[M,\partial_{\mu}M]U^{\dagger}  -\partial_{\mu}\alpha \, M+i [M,\partial_{\mu}M]  -Ui[M,\partial_{\mu}M] U^{\dagger} \nonumber  \\
&=&  C^{U}_{\mu}M+i[M,\partial_{\mu}M],  \eea
where
\be  C^{U}_{\mu}=C_{\mu}-  \,\partial_{\mu}\alpha,   \label{effU1tr}\ee
showing that $C_{\mu}$ indeed transform as a  $U(1)$ gauge field.

The SU($N$) transformation law is a little more subtle, as $E_{\mu}$ transforms nontrivially.
The  SU($N$) transformations commuting with $M$ are
\begin{equation}  U=  \exp(i\omega^{\textsf{A}}et^{\textsf{A}}\bar{e})=
e\, \Omega \, \bar{e}+z\bar{z}\;, \qquad  \Omega =   \exp(i\, \omega^{\textsf{A}}t^{\textsf{A}})
\;.   \end{equation}
where  $(t^{\textsf{A}})^{i}_{j}$  ($i,j =1,2,\ldots N$) are the standard SU($N$)  generators in the fundamental representation;   $\Omega^{i}_{j} \subset$  SU($N$) acts in the local (or dual)  space\footnote{This is analogous to the local Lorentz transformations in the Tetrad formalism of general relativity.}.

Again, the transformation properties of various terms
can be nicely summarized into
\begin{equation}i\partial_{\mu}UU^{\dagger}=ei\partial_{\mu}\Omega\Omega^{\dagger}\bar{e}+E_{\mu}-i[M,\partial_{\mu}M]  - UE_{\mu}U^{\dagger}  +Ui[M,\partial_{\mu}M]U^{\dagger}\end{equation}
so that the Cho connection (\ref{Ansatz}) is seen to transform as
\bqa    A &\to&   A^{U}_{\mu}=   U  ( A_{\mu} + i  \de_{\mu})  U^{\dagger} \non \\
 &  =&
C_{\mu}M+e\Omega B_{\mu}\Omega^{\dagger}\bar{e}  -  ei\partial_{\mu}\Omega\Omega^{\dagger}\bar{e}-E_{\mu}+i[M,\partial_{\mu}M]
 \non \\  &  =&
    C_{\mu}M+eB^{U}_{\mu}\bar{e}+i[M,\partial_{\mu}M]-E_{\mu} \eea
where
\begin{equation} B^{U}_{\mu}=\Omega B_{\mu}\Omega^{\dagger}-i\partial_{\mu}\Omega\Omega^{\dagger}.\end{equation}
 That is, the net effect of the original gauge transformation
 $ A_{\mu} \to A_{\mu}^{U}$
 is the gauge transformation of the  $B_{\mu}$ field in the dual space
 \be B_{\mu}  \to    B^{U}_{\mu}=\Omega B_{\mu}\Omega^{\dagger}-i\partial_{\mu}\Omega\Omega^{\dagger}\;, \qquad  M \to M, \qquad E_{\mu} \to E_{\mu.}  \label{localBtr}  \ee

It is interesting to note that  the  same transformation
$ B_{\mu}  - E_{\mu}  \to  B^{U}_{\mu} - E_{\mu} $
can be induced  simply by the local SU($N$)  transformation of the vielbein
\be  e(x)\rightarrow e(x)\Omega(x)\;. \label{vielbeins}   \ee
In fact, under this transformation
$M$ remains invariant while
\be  E_{\mu}\rightarrow E_{\mu}+ei\partial_{\mu}\Omega\Omega^{\dagger}\bar{e}, \qquad
eB_{\mu}\bar{e}\rightarrow e\Omega B_{\mu}\Omega^{\dagger}\bar{e}\;.    \ee
We note furthermore that, without the contribution of the vector  $E_{\mu}$,  the gauge field would not be form invariant, and that the above-mentioned equivalence between the original gauge transformations and the local dual SU($N$)  transformations would be lost.

The fact that the original SU($N$) $\subset$ SU($N+1$) gauge transformation can be expressed  {\it  either } as  the local SU($N$) gauge transformation of the $B_{\mu}$ field  Eq.~(\ref{localBtr})  {\it or} as  the local vielbein transformation, Eq.~(\ref{vielbeins}),    means also the following.  The gauge field $A_{\mu}$ written in terms of our decomposition Eq.~(\ref{Ansatz})  is {\it invariant} under the simultaneous local SU($N$) transformations
\be      e(x)\rightarrow e(x)\Omega(x)^{\dagger}, \qquad  B_{\mu}  \to    B^{U}_{\mu}=\Omega B_{\mu}\Omega^{\dagger}-i\partial_{\mu}\Omega\Omega^{\dagger}\;.
\label{simul}\ee
exhibiting the redundancy of the paramerization (\ref{Ansatz}), and explaining the counting Eq.~(\ref{counting}).

\subsection{Generalized Cho action} \label{azsez}



We have seen that  the local  U$(1) \times$ SU($N$) gauge transformations on $A_{\mu}$
\be  A \to   A^{U}_{\mu}=   U  ( A_{\mu} + i  \de_{\mu})  U^{\dagger}
\label{invloc} \ee
are equivalent to the local transformations
Eq.~(\ref{effU1tr}),   Eq.~(\ref{localBtr}) of $B_{\mu}$ and $C_{\mu}$ fields,
while $M$ remains invariant.
As the action
\be    \Tr \,  F_{\mu \nu} F^{\mu  \nu}, \qquad   F_{\mu \nu} =     \de_{\mu} A_{\nu} -   \de_{\nu} A_{\mu} - i [ A_{\mu}, A_{\nu}]
\ee
is by construction invariant under Eq.~(\ref{invloc}),  after the action has been calculated and simplified  the resulting expression must contain
only objects   {\it  invariant}  under the dual local transformations of Eq.~(\ref{effU1tr}) and  Eq.~(\ref{localBtr}).
Plugging (\ref{Ansatz}) into the Yang-Mills field-strength:
\be  F_{\mu \nu} =  \de_{\mu} A_{\nu} -   \de_{\nu} A_{\mu} - i  [ A_{\mu}, A_{\nu}]
\ee
we find indeed
\be\label{field}
F_{\mu\nu}=(C_{\mu\nu}+H_{\mu\nu})M + eB_{\mu\nu}\bar{e},
\ee
where
\be   C_{\mu\nu}\equiv\de_{\mu}C_\nu-\de_{\nu}C_\mu, \qquad  B_{\mu\nu}\equiv\de_{\mu}B_\nu-\de_{\nu}B_\mu-i [B_\mu,B_\nu] \ee
are respectively the abelian
and the SU($N$) field-strengths.
\be H_{\mu\nu}\equiv\Tr([\de_\mu M,\de_\nu M]M)  \ee
 is an analogue of the $H_{\mu\nu}$ field defined by Faddeev and Niemi, and describes the magnetic degrees of freedom.  Notice that the presence of the $E_{\mu}$ field is crucial to obtain such a simple formula for the field-strength: the terms proportional to $E_{\mu}$ combine
to cancel the component of $[\de_\mu M,\de_\nu M]$ orthogonal to $M$.

We can now straightforwardly evaluate the Lagrangian and find
\be
\mathcal{L}=\Tr F_{\mu\nu}F^{\mu\nu} = (C_{\mu\nu} + H_{\mu\nu})(C^{\mu\nu} + H^{\mu\nu}) + \tr B_{\mu\nu}B^{\mu\nu},  \label{ActionCho}
\ee
where $\tr$ indicates the trace over SU($N$). From (\ref{field}) we can straightforwardly derive Yang-Mills equations evaluating $D^{\mu}F_{\mu\nu}$.
We find the following two equations:
\be
\begin{aligned}
0=&\de^{\mu}(C_{\mu\nu}+H_{\mu\nu}),\\
0=&\de^{\mu}B_{\mu\nu}-i [B^\mu,B_{\mu\nu}].
\end{aligned}
\ee
We recognize these as the equations obtained varying $\mathcal{L}$ with respect to $C_{\mu}$
and $B_{\mu}$\footnote{The variation of $\mathcal{L}$ with respect to $M$, with an appropriate Lagrange multiplier, can be shown not to introduce any new equation.}. Generically, the Faddeev-Niemi equations are just a subset of the Yang-Mills ones, as noted in \cite{noi}.   However in this case the problem does not arise: the equations obtained varying the action with respect to the U($N$) connection imply the full SU($N+1$) YM equations.

\section{Non-Abelian Faddeev-Niemi decomposition for SU($3$) Yang Mills theory } \label{AlSez}

The action following from our Cho connection (\ref{ActionCho}) has a very simple interpretation. It consists of the U($1$) part describing the electric and magnetic contributions, and the SU($N$) action orthogonal to the U($1$), exhibiting the dynamical symmetry breaking ${\rm SU}(N+1)  \to   {\rm U}(N).$ Even though very elegant, such a
picture is clearly an oversimplification due to the reduction of degrees of freedom inherent in the Cho approach.
In the case of the SU($2$) theory with symmetry breaking ${\rm SU}(2)  \to   {\rm U}(1)$,  Faddeev and Niemi attempted to improve it by keeping $2$  (on-shell) or $4$ (off-shell) more degrees of freedom in $A_{\mu}^{a}$.  These extra degrees of freedom appear as one (or two)  complex scalar fields carrying the charge under the U($1$).  By averaging out the fluctuations of these ``matter fields''  first, Fadeev and Niemi managed to obtain an effective action describing the magnetic degrees of freedom (corresponding to our $M$ field), and showing that these low-energy theory possessed knotlike solitons. They have subsequently generalized their construction to the case of  ${\rm SU}(N)  \to   {\rm U}(1)^{N-1}$. They proposed that this kind of effective Abelian action may describe the confinement phase of QCD.

It is quite natural to ask if an analogous completion is possible in the non-Abelian scenario, ${\rm SU}(N+1)  \to   {\rm U}(N).$
Given formula (\ref{Ansatz}) one might think that just by adding a term of the form $e_{i}\Phi_{\mu }^{i}\bar{z}+h.c.$, where $\Phi^{i}_{\mu}$ contains a complex field in the fundamental of SU($N$), we can introduce the remaining 2N degrees of freedom, to complete the decomposition \`a la Fadeev-Niemi.   However, such a term clearly introduces more than  $2N$ degrees of freedom, unless the four vector components are all proportional.  Sticking to the idea of introducing precisely  $2N$ more fields, the only possibility would be to write $e_{i}\Phi^{i}\partial_{\mu}\bar{z}z\bar{z}+h.c.$, but such a term is not invariant under $z\rightarrow e^{i \alpha}z$:    $z$  would cease to be $2N$ dimensional coordinates of $\mathbb{CP}^{N}$.

It could be a more fertile idea to try to find a natural generalization  the ``off-shell decomposition'' found in \cite{FNvinc}  to our
non-Abelian breaking scenario,
rather than to insist on  generalizing the original  ``on-shell complete'' FN  decomposition \cite{FN1,FN2}.
 A possible such decomposition for SU($3$) Yang-Mills theory   has the form,
\be
A_{\mu} =    A_{\mu}^{Cho}  + \rho \, \partial_\mu M + \sigma_{I} (\partial_\mu N^{I} )^\perp     \label{cong}
\ee
where (Eq.~(\ref{Ansatz}))
\[  A_{\mu}^{Cho} =   C_{\mu}^{(0)}  \, M  +  i  [M, \de_{\mu} M] +  { B}_{\mu}  - E_{\mu}
\]   and
where
\be  N^{I} \equiv   e\,\tau^{I} \,  \bar{e}\;, \qquad  I=1,2,3
\ee
($\tau^{I}$  are the standard constant Pauli matrices). We  keep in (\ref{cong})  the part of  $\de_{\mu}N^{I}$  orthogonal  to U(2),
\be   (\de_{\mu}N^{I})^{\perp} \equiv  z {\bar z} \, \partial_\mu e \, \tau^{I} \bar e +  h.c.  =    [[ \de_{\mu} N^{I}, M], M]\;,
\ee
(recalling   $z{\bar z}+ \sum  e \, {\bar e} = {\mathbf 1}$)    to avoid mixing with $B_{\mu}$.
The first four terms already appear in (\ref{Ansatz}), the matrices $N^{I}$ are the Pauli matrices embedded in SU($3$)  whereas $\rho$ and $\sigma_{I}$ are real scalars.  This formula contains 12 on-shell degrees of freedom in the Cho sector and four parameters describing ``matter fields''.  This might appear to be the right number of on-shell degrees of freedom (16) to describe the original Yang-Mills theory.

Unfortunately,  (\ref{cong}) is not form invariant under U($2$) rotations (even under the ${\rm U}(1)$ rotations about the $M$ axis the $\rho$ and $\sigma$ fields pick up a phase). This means that by acting with  U($2$) gauge transformations we generate new terms, which must be taken into account in the spirit of the original Faddeev-Niemi construction.  However, as we will now see, the form invariance can be achieved by simply promoting $\rho$ and $\sigma^i$ to complex fields, and by adding the necessary Hermitian conjugate terms.

Using the $e_{i}$ and $z$ variables we can rewrite the part of formula (\ref{cong}) proportional to the $\rho$ and $\sigma_{i}$ fields as
\be\label{real}
[\rho \, \partial_\mu \, M + \sigma_{I}  (\partial_\mu N^{I})^\perp]^{a}_{b} =      e^{a}   H  {\bar e}_{c}\, \de_{\mu}z^{c}\bar{z}_{b} + h.c.\;,
\ee
    where
\be      H^{j}_{k} \equiv   \rho\, \delta^{j}_{k}-\sigma_{I}\,  (\tau^{I})^{j}_{k}\,.
\ee
The 2x2 matrix $H$  (with indices $i,j,..$ of the dual space) is Hermitian if the $\rho$ and $\sigma^i$ fields are real.  It is easy to see that the action of a
${\rm U}(2)$ gauge transformation (\ref{cong})  produces the transformation
\be H\rightarrow \Omega  \, H \label{twodoublets} \ee
 (where $\Omega $ is the induced U($2$) transformation  in the dual space, see Subsection~\ref{UNCho}).
According to the polar decomposition, every complex matrix $\Phi$ can be written in the form $\Phi=UH$, where $U$ is unitary and $H$ Hermitian; applying this result to
our case, we deduce that by acting linearly with the U($2$) unbroken gauge group on $H$ we generate the general complex matrix $\Phi$.  Thus, the requirement of form invariance under U(2) rotations
leads us to replace (\ref{real}) with
\be\label{complex}
e^{a} \Phi \, {\bar e}_{c}\, \de_{\mu} z^{c}\bar{z}_{b} + h.c.  =  e^{a}_{i}   \Phi^{i}_{j}  {\bar e}_{c}^{j}  \,  \de_{\mu} z^{c}\bar{z}_{b} + h.c.,   \qquad   \Phi^{i}_{j} \equiv  \rho\, \delta^{i}_{j}-\sigma_{I} (\tau^{I})^{i}_{j},  \ee
where now  $\rho(x)$ and $\sigma_{I}(x)$  ($I=1,2,3$)  are complex fields.

Most interestingly, the transformation law  $\Phi \rightarrow \Omega  \, \Phi   $  means that
\be \Phi  =  \left( \begin{array}{c}{\bf \phi}_1^{(1)} \\{\bf \phi}_2^{(1)}\end{array}  \begin{array}{c}{\bf \phi}_1^{(2)} \\{\bf \phi}_2^{(2)}\end{array} \right)\ee transform as two doublets of  SU($2$), both with unit U($1$) charge.
To summarize, then, our Ansatz (\ref{cong}) with complexified  ($\rho$, $\sigma_{I}$),
\be    A_\mu =   A_{\mu}^{Cho}  + e^{a} \Phi\, {\bar e}_{c}\, \de_{\mu} z^{c}\bar{z}_{b} + h.c.,
\ee
  represent the following degrees of freedom:
$C_{\mu}$ ($4$) and $B_{\mu}$ ($12$)  are   U($2$) gauge fields,  $M$ ($4$) is the $\mathbb{CP}^{2}$ coordinates of the embedding  $ {\rm U}(2) \hookrightarrow  {\rm SU}(3)$,
and two complex ``matter''  fields both of which are doublets of   SU($2$)   ($8$).    Altogether they represent precisely  $28= 32-4$ (off-shell) degrees of freedom, corresponding to the partial gauge fixing of the original SU($3$) theory.

By introducing
 $\rho\equiv\rho_1+i \rho_2$ and $\sigma_{I}\equiv \alpha_{I}+i  \beta_{I}$
our connection can be written in an alternative form
\be\label{inv}
A_\mu=A_{\mu}^{Cho}+\rho_1\de_\mu M-i\rho_2 [\de_\mu M,M]+\alpha_{I}\,(\de_\mu N^{I})^\perp-i\,\beta_{I} [\de_\mu N^{I}, M]\;.
\ee
In this form we recognize the pair  $\rho_1, \rho_{2}$ as analogues of  ($\rho, \sigma$) fields introduced in the original Faddeev-Niemi decomposition \cite{FN1}.

\section{Monopoles and Obstructions} \label{topsez}

Our generalized Cho-Faddeev-Niemi connection describes very naturally magnetic monopoles and instantons, as the abelian decomposition considered by Faddeev and Niemi.
We discuss here some of the specific issues arising in our nonabelian scenario.


\subsection{Topologically stable configurations}

The Higgs field $M$ defines a function from spacetime to $\mathbb{CP}^N$. Stationary solutions in $(3+1)$-dimensional spacetime provide a map from $\mathbf{R}^3$ to $\mathbb{CP}^N$.  Configurations which are topologically trivial at infinity may be characterized by maps in which spatial infinity is compactified, so these are maps from $S^3$ to $\mathbb{CP}^N$.  Topologically stable Higgs fields correspond to maps which cannot be smoothly  deformed to constant maps.  These are characterized by the homotopy group
\be
\pi_3(\mathbb{CP}^N)=\left\{\begin{array}{cl} \mathbb{Z} \rm{\ \ \ when\  \mathit{N}=1}\\ 0 \rm{\ \ \ when\  \mathit{N}>1.}\\\end{array}\right.
\ee

In Refs.~\cite{nodi1,nodi2} the authors were interested in the case $N=1$, in which this homotopy group was nontrivial, and so there were topologically nontrivial Higgs fields which were trivial at infinity.  The authors claimed, after a numerical analysis, that such topologically nontrivial fields correspond to actual solutions to the equations of motion, which they identified with various knots.  They later extended their analysis to other symmetry breaking patterns ending with abelian groups, in which the Higgs field is valued in a flag manifold with nontrivial $\pi_3$ and so such knots continue to exist.

In our case $N>1$, however,  and so there is no topologically nontrivial Higgs field configuration which is topologically trivial at infinity.  There are no topologically stable knot solutions, analogous to those of Faddeev-Niemi.

Topologically nontrivial Higgs fields nonetheless exist in our case as well.  There are characterized by the topology of the Higgs field on the 2-sphere at spatial infinity.  Stable equivalences classes of such Higgs fields carry charges in
\be
\pi_2(\mathbb{CP}^N)=\mathbb{Z}, \qquad  \forall N\;.
\ee
This group is nontrivial in our case as well.  A configuration representing the element $q$ of this homotopy group is a Wu-Yang ('t Hooft-Polyakov in the presence of elementary Higgs fields) monopole with $q$ units of magnetic charge.  While such configurations exist with symmetry breaking SU($N+1)\rightarrow$ U($N$), the SU matrices  cannot be continuously defined on the entire spacetime in these cases.
  We will now describe this topological obstruction.

\subsection{A minimal monopole} \label{wy}

If a gauge could be chosen so that
\be   z = \left(\begin{array}{c} 1 \\  0  \\\vdots \\ 0\end{array}\right)   \label{everywhere}\ee
everywhere,  then  the remaining gauge freedom together with an SU($N$) acting on $i$  (acting on different $e_{i}$'s  as a ``dual''  group),   could be used to set
\be    e_{i}^{a}  = \delta_{i}^{a-1},  \qquad   e_{i}^{1}=0\;.  \label{everywhereBis}
\ee
\[  \qquad i = 1,2,\ldots, N, \quad  a = 2,3, \ldots, N+1\;.
\]
In  other words, in that gauge
\be    U =  {\mathbbm 1}_{N+1 \times N+1}\;, \qquad  \sum  e\, {\bar e} =  \left(\begin{array}{cc}0 &  \\    &  {\mathbbm 1}_{N \times N}\end{array}\right)
\ee
and   $B_{\mu}$ represents simply the original gauge fields in  SU($N$)  $\subset$  SU($N+1$).
The problem is that for general gauge field configurations it is not possible to choose the gauge so that Eqs.~(\ref{everywhere}), (\ref{everywhereBis})  hold
everywhere.  For instance take a ``spin $1/2$'' wave function  embedded in an  SU(2) subgroup:
\be   z = \left(\begin{array}{c} e^{-i \varphi/2}  \cos \frac{\theta}{2}   \\  e^{i \varphi/2}  \sin \frac{\theta}{2}     \\  0 \\  \vdots \\ 0\end{array}\right)  \;; \label{monopole}\ee
then
\be     M =     \frac{1}{2}   \left(\begin{array}{ccc}{\bf n}\cdot \tau  &  &  \\ & 0 &  \\ &  & \ddots\end{array}\right)
+    \frac{1}{2}   \left(\begin{array}{ccc}{\mathbbm 1}_{2\times 2}  &  &  \\ & 0 &  \\ &  & \ddots\end{array}\right)
- \frac{1}{N+1} {\mathbbm 1}\;;
 \label{Mmonop}
\ee
\be     {\bf n}\cdot \tau =  \left(\begin{array}{cc}\cos \theta  & e^{-i\varphi } \sin \theta  \\   e^{i \varphi} \sin \theta & -\cos \theta \end{array}\right) =
 \frac{\bf r}{r}  \cdot \tau
\ee
The second term  of Eq.~(\ref{Ansatz}), $i \,  [M, \de_{\mu} M] $,    gives then
\be     \frac{1}{4}  \,    \de_{\mu} {\bf n}  \times {\bf n}
\ee
lying in SU$(2)\subset$ SU($N+1$), which is precisely the singular Wu-Yang monopole solution \cite{Wuyang}.  As  in the original Faddeev-Niemi decomposition, it is quite possible that the presence of other fields $\rho, \sigma_{I}$ smears the singularity \cite{FN1,KonTak};  there is nothing that prohibits that the collection of these regularized monopole configurations dominates the infrared limit of the Yang-Mills theory.

 Clearly, we can build other monopole solutions by
  taking $z$ of the form
\be   z = \left(\begin{array}{c} e^{-i \varphi/2} \cos \frac{\theta}{2}   \\  0 \\  \vdots \\ 0   \\  e^{i \varphi/2}  \sin \frac{\theta}{2}     \\  0 \\  \vdots \\ 0\end{array}\right)  \;, \label{Nmonopole}\ee
where only the first and i-th entries are nontrivial; this corresponds to the Wu-Yang monopole embedded in the plane (1,i). If we now calculate the field-strength tensor
using the connection just defined we obtain
\be
F_{ij}=\epsilon_{ijk}\frac{r_k}{r^3}M.
\ee
We thus construct N monopole solutions, differing only in the embedding of SU($2$) into SU($N+1$), which form a multiplet in the fundamental representation of $\widetilde{{\rm SU}(N)}$; this is precisely the nonabelian GNO monopole (singular at the origin).

Can one define an unbroken SU($N$) group  orthogonal to the U(1) defined by the direction of $z$ ($M$) globally?
This is of course the ``topological obstruction''  noted by Aboulsaad et. al. \cite{CDyons}.
The transformation $U$  that would bring (\ref{Mmonop})  into  the fixed U(1)  generator  $T^{(0)}$   (Eq.~(\ref{MtoU}),  Eq.~(\ref{UtoM}))  in this case   is
\be      U =    \left(\begin{array}{cc}V & 0 \\0 & {\mathbbm 1}_{N-1 \times N-1}\end{array}\right), \qquad V=    \left(\begin{array}{cc}e^{-i \varphi/2}  \cos \frac{\theta}{2}  & e^{-i \varphi/2}  \sin \frac{\theta}{2}   \\e^{i \varphi/2}  \sin \frac{\theta}{2}   & -e^{i \varphi/2}  \cos \frac{\theta}{2}  \end{array}\right)
\ee
 The generators of SU($N$),     $T^{(A)}$, $A=1,2,\ldots N^{2}-1$,
\be     \left(\begin{array}{cc}0 &  \\ & T^{(A)}\end{array}\right)
\ee
which involve the second column or the second row,  such as  SU(2) generators
\be    \left(\begin{array}{cccc}0 &  &  &  \\ & \tau^{1,2} &  &  \\ &  & 0 &  \\ &  &  & \ddots\end{array}\right)\ee
do not have well-defined set of images
\be    U\,   \left(\begin{array}{cc}0 &  \\ & T^{(A)}\end{array}\right) \, U^{\dagger}\;,  \label{image}
\ee
as is well known \cite{CDyons}.


\subsection{The obstruction in general}

More generally, recall that a Higgs field $M$ which breaks SU($N+1$) to U($N$) defines an element of the projective space $\mathbb{CP}^N$.  Given a choice of $M$, the matrix $U$ defines an embedding of U($N$) into SU($N+1$) such that the U($N$) commutes with $M$.  For any given value of $M$, there are many such embeddings.  Different points in spacetime, in general, have different values of the Higgs field $M$, and so different embeddings.  Thus in principle one would like to define $U$ as a function of $M$, so that an embedding exists at each point $x$ corresponding to the value $M(x)$.

However no continuous function ${\cal U}$ of $M$ exists such that ${\cal U}$ defines an embedding of U($N$) that commutes with $M$ for all $M$.  In fact, as U($N$) is nonabelian, it is not even possible to define its generators for all values of $M$, and therefore it is not possible to define the action of a global U($N$) symmetry in the presence of an arbitrary Higgs field configuration.  

On the other hand, such a function does exist if one restricts $M$ to lie in a particular subset of $\mathbb{CP}^N$, for example if one removes a $\mathbb{CP}^{N-2}$ then the SU($N$) subgroup can be globally defined.  The largest such subset is $\mathbb{CP}^N$ with a $\mathbb{CP}^{N-1}$ removed.  One may remove any $\mathbb{CP}^{N-1}$ which represents the element $[1]$ of the $(2N-2)$nd homology group of $\mathbb{CP}^N$
\be
[\mathbb{CP}^{N-1}]=1\in\rm{H}_{2N-2}(\mathbb{CP}^N)=\mathbb{Z}. \label{om}
\ee
The points in spacetime in which the Higgs field has values in the removed $\mathbb{CP}^{N-1}$ correspond to a Dirac string, on which an embedding of U($N$) does not exist.  One may choose any $\mathbb{CP}^{N-1}$ satisfying (\ref{om}), different choices give different locations for the Dirac string.  Thus, as in the case of the conventional Dirac string, the location of the string is a gauge choice.  However, as there is no U($N$) embedding on the Dirac string, U($N$) global symmetries cannot be defined in such a gauge.

't Hooft-Polyakov monopoles have nontrivial Higgs field profiles.  In particular, the Higgs field at large distances from the monopole is topologically nontrivial.  It varies with respect to the 2-sphere at spatial infinity, sweeping out
\be
[S^2]=1\in\rm{H}_{2}(\mathbb{CP}^N)=\mathbb{Z}. \label{om2}
\ee
A charge $q$ monopole represents the element $q$ in this second homology group.
The key topological fact is that the intersection product between the space of obstructed values of the Higgs field $\rm{H}_{2N-2}(\mathbb{CP}^N)=\mathbb{Z}$ and the space of Higgs field values at infinity in a monopole configuration $\rm{H}_{2}(\mathbb{CP}^N)=\mathbb{Z}$ is nontrivial.  In fact, the intersection of the cycles in Eqs.~(\ref{om}) and (\ref{om2}) generates $\rm{H}_{0}(\mathbb{CP}^N)=\mathbb{Z}.$  This means that the Higgs field in a charge $q$ monopole configuration necessarily intersects the obstruction surface (\ref{om}) with multiplicity $q$.  Therefore if there is a net monopole charge, there will always be a topological obstruction to a global embedding of U($N$) in SU($N+1$).  

Recall that there is no obstruction when $M$ is valued in $\mathbb{CP}^N$ with $\mathbb{CP}^{N-1}$ removed.  One may describe this space using homogeneous coordinates $w=(1,v_1,...,v_N)$ which are related to $z$ by division by $z^1$.  In other words
\be
v_k=\frac{z^{k+1}}{z^1}. \label{v}
\ee
The $\mathbb{CP}^{N-1}$ which is removed corresponds to the points $z^1=0$, where (\ref{v}) is ill-defined.

Now $M$ and $U$ (Eq.~(\ref{UtoM})) may be expressed globally in terms of $w$.  $M$ is given by
\be
M=   \frac{1}{|w|^2}  w\overline{w}-\frac{1}{N+1}\mathbf{1}
\ee
and the corresponding $U$ can be given explicitly (\ref{Noobst}).  The image of ${\rm U}(N)\subset {\rm SU}(N+1)$  (Eq.~(\ref{image})) is then everywhere well defined in terms of   $v_{i}$,  Eq.~(\ref{explicitU}), Eq.~(\ref{explicitUBis}).  Of course, we lose also monopoles by restricting to  configurations with $\mathbb{CP}^{N-1}$ removed:   this is just another way of saying that the presence of monopoles in nonabelian scenario necessarily leads to the topological obstruction, if the full global U($N$) symmetry is to be retained.


\subsection{Instantons and merons in the SU($3$) theory}

We have seen that our Cho connection (\ref{Ansatz}) can be used to describe nonabelian magnetic monopoles. Obviously the same holds for our decomposition of the SU($3$) YM connection (\ref{inv}), since we recover the Cho conection just setting the $\rho$ and $\sigma_i$ fields to zero. However, these extra fields allow us to recover many other configurations of physical interest, confirming the validity of our decomposition, as we will see in this section.

In Ref.~\cite{FN2} the authors observe that Witten's Ansatz for instantons is contained in their decomposed SU($2$) connection; the analogous statement holds in our case.
Let us recall formula (\ref{inv}):
$$A_\mu=A_{\mu}^{Cho}+\rho_1\de_\mu M-i \rho_2 [\de_\mu M,M]+\alpha^i(\de_\mu N_i)^\perp-i \beta^i [\de_\mu N_i,M].$$
As well known, the SU($N$) instantons are obtained by embedding the standard instanton in various SU($2$) subgroups, either in  the unbroken SU($2$) (whose generators commute with M), or in  broken SU($2$)'s  such as  U-spin or V-spin. The instanton in the unbroken subgroup can be recovered simply choosing the appropriate $B_{\mu\, j}^{i}$, so there is nothing new to add. If we take $z$ of the form (\ref{Nmonopole}),  we obtain the analogue of the {\bf{n}} vector of (\cite{FN1}) for the U-spin or V-spin.  By setting
\be
C_i=Cx^i,\qquad B_{\mu}^{i}=\alpha_i=\beta_i=0,
\ee
with $C$, $C_t$, $\rho_1$ and $\rho_2$ depending on $t$ and $r$ only, we obtain exactly Witten's Ansatz for instantons embedded in the "broken'' SU($2$) subgroups. With the same technique, it is straightforward to recover also meron-like configurations: taking $z$ as before and setting
\be
1+\rho_2=r\frac{\de}{\de r}R(x),\qquad C_{\mu}=B_{\mu}^{i}=\rho_1=\alpha_i=\beta_i=0,
\ee
we obtain a single meron or a multi-meron configuration, depending on the specific form of the function $R(x)$.
A similar result could be obtained for the general SU($N$) case if one can find a ${\rm U}(N)$ invariant decomposition which contains the terms $\rho\, \de_\mu M-i \sigma [\de_\mu M,M]$. We have not found such a solution which does not introduce too many parameters.

\section{Concluding remarks}

In spite of subtleties discovered recently \cite{noi,Niemi} the Faddeev-Niemi decomposition \cite{FN1,FN2} may still be a useful tool for analyzing the infrared behavior of some strongly-interacting gauge theories such as QCD.  Motivated by the lack of clear evidence of dynamical abelianization in real-world QCD, in this paper we have studied  the structure of a possible nonabelian Faddeev-Niemi decomposition corresponding to dynamical gauge symmetry breaking ${\rm SU}(N+1)\to {\rm U}(N)$.  The generalized Cho connection has been constructed, and its dual gauge symmetry transformation properties have been clarified.

It turns out that it is not possible to straightforwardly generalize the ``on-shell'' FN decomposition to the nonabelian
scenario.  Restricting to the ${\rm SU}(3)$ theory, however, we have found a very natural ``off-shell'' FN decomposition of nonabelian type, corresponding to the partial gauge fixing ${\rm SU}(3)/{\rm U}(2)$, with the correct number of degrees of freedom.   We find it interesting that in this decomposition, the effective degrees of freedom correspond to those appearing in the Cho connection (the $\mathbb{CP}^
{2}$ degrees of freedom describing the embedding  $ {\rm U}(2) \hookrightarrow  {\rm SU}(3)$ and the
 ${\rm U}(2) $ gauge-field degrees of freedom)  plus  two complex scalars in the doublet representation of  ${\rm U}(2)$.
 
A consequence of the minimal breaking ${\rm SU}(3)\to {\rm U}(2)$  is that the analogue of the FN knot solitons \cite{FN1,FN2, nodi1,nodi2}
does not exist in this case, in contrast to the dynamically abelianized system, ${\rm SU}(3)\to {\rm U}(1)^{2}$ .

Another decomposition for the ${\rm SU}(3)$ Yang-Mills fields, by using the orthogonal Zweibeins in the planes perpendicular to the 
Cartan subalgebra directions  (which has some formal similarity to ours), has been studied by Bolokhov and Faddeev \cite{FNZwei}.  The relation between their work and ours is not very clear to us, but we are under the impression that their work basically hinges upon the idea of dynamical abelianization and is in that sense quite distinct from the present work.
 
Of course, no change of variables can eliminate in itself the problems \cite{CDyons,DFHK} associated with the nonabelian monopoles: we have indeed
discussed one of them (topological obstruction) in our more general context here. It is quite possible that our nonabelian symmetry breaking cannot be dynamically realized
in the infrared in the pure ${\rm SU}(3)$ Yang Mills theory.  Whether or not the nonabelian scenario  ${\rm SU}(3)\to {\rm U}(2)$ is relevant in the infrared in theories with light quarks, is a purely dynamical question which no classical argument such as ours can answer.  Results from supersymmetric models  strongly suggest that a certain amount  of light matter fields are essential to prevent nonabelian monopoles from becoming too strongly coupled in the infrared \cite{Duality}. The coupling of our nonabelian FN variables to the quarks and its possible consequences are left for future study.

\section* {Acknowledgment}   We thank Daniele Dorigoni for useful discussions.  J.E. is supported by the CAS Fellowship for Young International Scientists grant number 2010Y2JA01.


\appendix

\section{U($N$) gauge transformations on the Cho connection } \label{transsez}

In this Appendix we show in detail how the U($N$) gauge transformations act on the generalized Cho connection and check its form invariance.

The U(1) transformation around the $M$ direction acts by definition as
\be  U =  e^{i \alpha M}  =    e^{i \alpha  \frac{N}{N+1} }    z {\bar z}   +   e^{- i \alpha  \frac{1}{N+1} }    e {\bar e}\;.
\ee
where a convenient form
$     M=  ({N}/{N+1}) \, z {\bar z} -  ({1}/{N+1})\,   e {\bar e},
$
has been used.   The fields  $B_{\mu}$ and   $E_{\mu}$ are  easily seen to be invariant under such ${\rm U}(1)$ transformations.

$M$ field is invariant by definition;   a straightforward calculation leads to
\begin{equation}U \, \partial_{\mu}M \, U^{\dagger}=   \cos \alpha\,\partial_{\mu}M +\sin  \alpha\, i[M,\partial_{\mu}M] \;, \label{rot1} \end{equation}
and
\be  U i[M,\partial_{\mu}M]U^{\dagger}=i ^{2}  e^{-i\alpha  } e\, \partial_{\mu}\bar{e}\, z \, \bar{z} + \textrm{h.c.} =
\cos\alpha   \,  i  [M,\partial_{\mu}M]- \sin\alpha \, \partial_{\mu}M   \;,   \label{rot2}   \ee
in close analogy with the abelian decomposition discussed by Faddeev and Niemi \cite{FN1}.
The inhomogeneous term arising from the gauge transformation can then be written as
\begin{equation}i\partial_{\mu}UU^{\dagger}=\partial_{\mu}\alpha \, M-i[M,\partial_{\mu}M]  +Ui[M,\partial_{\mu}M]U^{\dagger}  \;. \end{equation}
The Cho connection (\ref{Ansatz}) can be easily seen to transform (leaving out the invariant terms $B_{\mu}-E_{\mu}$) as
\bqa  A_{\mu}  &  \rightarrow  & A^{U}_{\mu}\equiv UA_{\mu}U^{\dagger}-i\partial_{\mu}UU^{\dagger}    \nonumber  \\
&=&  C_{\mu}M+Ui[M,\partial_{\mu}M]U^{\dagger}  -\partial_{\mu}\alpha M+i [M,\partial_{\mu}M]  -Ui[M,\partial_{\mu}M] U^{\dagger} \nonumber  \\
&=&  C^{U}_{\mu}M+i[M,\partial_{\mu}M],  \eea
where
\be  C^{U}_{\mu}=C_{\mu}-  \,\partial_{\mu}\alpha,   \label{effU1}\ee
showing that $C_{\mu}$ indeed transform as a  $U(1)$ gauge field.

The  SU($N$) transformations commuting with $M$ can be written   by exponentiating
\[ e\, t^{\textsf{A}} \, \bar{e}, \qquad [ e\, t^{\textsf{A}} \, \bar{e}, M]=0 \;,     \]
that is
\begin{equation}  U=  \exp(i\omega^{\textsf{A}}et^{\textsf{A}}\bar{e})=
e\, \Omega \, \bar{e}+z\bar{z}\;, \qquad  \Omega =   \exp(i\omega^{\textsf{A}}t^{\textsf{A}})
\;.   \end{equation}
where  $(t^{\textsf{A}})^{i}_{j}$  ($i,j =1,2,\ldots N$) are the standard SU($N$)  generators in the fundamental representation;   $\Omega^{i}_{j} \subset$  SU($N$) acts in the local (or dual)  space\footnote{This is analogous to the local Lorentz transformations in the Tetrad formalism of general relativity.}.  Note that
\[  U^{\dagger} U= {\mathbbm 1}; \qquad  U\, M \, U^{\dagger} = M.
\]

One finds
\begin{equation}U\partial_{\mu}MU^{\dagger}=-(e\Omega\partial_{\mu}\bar{e}z\bar{z}+z\bar{z}\partial_{\mu}e\Omega^{\dagger}\bar{e})  \label{deM}\end{equation}
while
$$Ui[M,\partial_{\mu}M]U^{\dagger}=i[M,U\partial_{\mu}MU^{\dagger}]=-i[e\bar{e},U\partial_{\mu}MU^{\dagger}]$$
By using (\ref{deM})  one finds
\begin{equation}Ui[M,\partial_{\mu}M]U^{\dagger}=ie\Omega\partial_{\mu}\bar{e}z\bar{z}-iz\bar{z}\partial_{\mu}e\Omega^{\dagger}\bar{e}.
\label{deMM}\end{equation}
Note that under the SU($N$) transformation  the vector $E_{\mu}$  is not invariant: it transforms as
\begin{displaymath}UE_{\mu}U^{\dagger}=iUe(\bar{e}\partial_{\mu}e-\tfrac{\mathbb{I}}{N}\textrm{Tr}[\bar{e}\partial_{\mu}e])\bar{e}U^{\dagger}=\end{displaymath}
\begin{equation}ie\Omega(\bar{e}\partial_{\mu}e-\tfrac{\mathbb{I}}{N}\textrm{Tr}[\bar{e}\partial_{\mu}e])\Omega^{\dagger}\bar{e}=ie\Omega\bar{e}\partial_{\mu}e\Omega^{\dagger}\bar{e}-e\tfrac{\mathbb{I}}{N}\textrm{Tr}[\bar{e}\partial_{\mu}e]\bar{e}\end{equation}
Also in this case we need to find out the inhomogeneous term
 $i\partial_{\mu}UU^{\dagger}$
 $$\partial_{\mu}U=e\partial_{\mu}\Omega\bar{e}+\partial_{\mu}e\Omega\bar{e}+e\Omega\partial_{\mu}\bar{e}-\partial_{\mu}e\bar{e}-e\partial_{\mu}\bar{e}$$
$$i\partial_{\mu}UU^{\dagger}=ei\partial_{\mu}\Omega\Omega^{\dagger}\bar{e}+i\partial_{\mu}e\bar{e}+ie\Omega\partial_{\mu}\bar{e}e\Omega^{\dagger}\bar{e}-i(1-e\bar{e})\partial_{\mu}\Omega^{\dagger}\bar{e}+ie\Omega\partial_{\mu}\bar{e}z\bar{z}+$$
$$+i\partial_{\mu}e\bar{e}-ie\partial_{\mu}\bar{e}-ie\partial_{\mu}\bar{e}e\bar{e}$$
which can be written, by using the transformation properties of
$E_{\mu}$ and    $i[M,\partial_{\mu}M]$, as
\begin{equation}i\partial_{\mu}UU^{\dagger}=ei\partial_{\mu}\Omega\Omega^{\dagger}\bar{e}+E_{\mu}-i[M,\partial_{\mu}M]  - UE_{\mu}U^{\dagger}  +Ui[M,\partial_{\mu}M]U^{\dagger}\end{equation}
From this we see 
that the $E_{\mu}$ term is crucial to obtain form invariance under SU($N$) as well;    it is seen that besides the two terms of the abelian case we must add two more terms that commute with the abelian transformation in order to absorb the term
 $-i\partial_{\mu}UU^{\dagger}$:
\begin{equation}A_{\mu}=C_{\mu}M + i[M,\partial_{\mu}M]  +eB_{\mu}\bar{e} -  E_{\mu}\label{decom2}  \end{equation}
where  $B_{\mu}$ is a vector field taking values  in the algebra of  SU($N$). The previous formula is exactly our generalized Cho connection (\ref{Ansatz}).
Thus
\bqa  &  A \to   A^{U}_{\mu}=   U  ( A_{\mu} + i  \de_{\mu})  U^{\dagger} \non \\
  &=   C_{\mu}M+e\Omega B_{\mu}\Omega^{\dagger}\bar{e}+Ui[M,\partial_{\mu}M]U^{\dagger}-UE_{\mu}U^{\dagger}   \non \\
& -  ei\partial_{\mu}\Omega\Omega^{\dagger}\bar{e}-E_{\mu}+i[M,\partial_{\mu}M]+UE_{\mu}U^{\dagger} -  Ui[M,\partial_{\mu}M]U^{\dagger}  \non \\  &  =
C_{\mu}M+e\Omega B_{\mu}\Omega^{\dagger}\bar{e}  -  ei\partial_{\mu}\Omega\Omega^{\dagger}\bar{e}-E_{\mu}+i[M,\partial_{\mu}M]
 \non \\  &  =
    C_{\mu}M+eB^{U}_{\mu}\bar{e}+i[M,\partial_{\mu}M]-E_{\mu} \eea
where
\begin{equation} B^{U}_{\mu}=\Omega B_{\mu}\Omega^{\dagger}-i\partial_{\mu}\Omega\Omega^{\dagger}.\end{equation}
 That is, the net effect of the original gauge transformation
 $ A_{\mu} \to A_{\mu}^{U}$
 is the gauge transformation of the  $B_{\mu}$ field in the dual  space
 \be B_{\mu}  \to    B^{U}_{\mu}=\Omega B_{\mu}\Omega^{\dagger}-i\partial_{\mu}\Omega\Omega^{\dagger}\;, \qquad  M \to M, \qquad E_{\mu} \to E_{\mu.}  \label{localB}  \ee

On the other hand,  this same transformation
\[  B_{\mu}  - E_{\mu}  \to  B^{U}_{\mu} - E_{\mu}
\]
can be induced  simply by the local SU($N$)  transformation of the vielbein
\be  e(x)\rightarrow e(x)\Omega(x)\;. \label{vielbein}   \ee
In fact, under this transformation
$M$ remains invariant while
\be  E_{\mu}\rightarrow E_{\mu}+ei\partial_{\mu}\Omega\Omega^{\dagger}\bar{e}, \qquad
eB_{\mu}\bar{e}\rightarrow e\Omega B_{\mu}\Omega^{\dagger}\bar{e}\;.    \ee

Note that such a conversion of the original gauge transformations into  the local ``flavor'' transformations  is quite analogous to the local Lorentz
transformations in the tetrad formalism of the general relativity.
Indeed by making a transformation    $e(x)\rightarrow e^{i\alpha(x)} e(x)$
the field $A_{\mu}$ remains invariant under a U(1) gauge transformation.

Moreover we note that without the contribution of the vector  $E_{\mu}$ the gauge field would not be form invariant, and that the above-mentioned equivalence between the original gauge transformations and the local SU($N$) flavor transformations would be lost.

\section{The Cho connection for  more general patterns of symmetry breaking \label{genecho}}

In Section~\ref{sec:Cho} we have considered the minimal breaking ${\rm SU}(N+1)\rightarrow {\rm U}(N)$, motivated by the idea that nonabelian monopoles might play a key role in  the infrared dynamics of QCD. In the context of some strongly interacting systems, it could be of interest to extend our construction of the Cho connection to  a more general pattern of symmetry breaking,  involving more than one surviving nonabelian factors as in ,
${\rm SU}(5) \to {\rm SU}(3)\times {\rm SU}(2)\times {\rm U}(1)$.

Our starting point is the construction of a parametrization for the coset $G/H$. In (\ref{Chosez}) we introduced the variable $z\in\mathbb{CP}^{N}$ in order
to construct the matrix $M$; as we have seen, the phase of $z$ is not observable and just the combination $z\bar{z}$, which is the projector on the one dimensional
space spanned by $z$, is physically meaningful. When the unbroken subgroup contains two nonabelian factors (e.g. ${\rm SU}(N+1-K)\times {\rm SU}(K)$), in order to say how
it is embedded in ${\rm SU}(N+1)$, it is enough to identify the K-dimensional subspace of $\mathbb{C}^{N+1}$ on which the ${\rm SU}(K)$ factor acts nontrivially (as is well known, the coset space in this case is the Grassmannian).
Equivalently, we just need to build the projector on this subspace. In order to do that, we introduce K orthonormal vectors $u_1,\dots,u_k$ and define a $N+1\times K$ matrix which is the generalization of our $z$ vector:
\be\label{ZK}
Z_K=  \left(    \left(\begin{array}{c}   \\ u_1  \\  \\  \end{array}\right)   \cdots   \left(\begin{array}{c}   \\ u_K  \\  \\  \end{array}\right)   \right)= U \left(\begin{array}{c}
0_{(N+1-K)\times K} \\
I_{K\times K}
 \end{array}\right).
\ee
In the previous formula U is a matrix in ${\rm SU}(N+1)$. The $(N+1)\times (N+1)$ matrix $Z_K Z_{K}^{\dagger}$ has obviously rank K and by means of a straightforward calculation one can see that it is a projector. It transforms in the adjoint of the unitary group (as can be inferred by the transformation property of $Z_K$) and its stabilizer is given by the block-diagonal unitary transformations, so its orbit has exactly the dimension of the Grassmannian. We can also rewrite the matrix $Z_K$ in the form
\be Z_K= \left(\begin{array}{c}
E \\
\sqrt{I_{K\times K}- E^{\dagger}E}
 \end{array}\right),\ee
which is the customary parametrization of the Grassmannian, used in the study of sigma models (see for instance \cite{Sigma}).

In order to construct the Cho connection for this pattern of symmetry breaking we introduce the traceless hemitian matrix
\be\label{MK}
M_K = Z_K Z_{K}^{\dagger}-\frac{K}{N+1}I_{N+1\times N+1},
\ee
which commutes with the whole residual gauge group and we complete the set $u_1,\dots,u_k$ to an orthonormal basis adding the vectors (in analogy with the case of minimal breaking)
\be
e_{i}^{a} \qquad a=1,\dots, N+1;\quad i=1,\dots, N+1-K.
\ee
We can thus accomodate in our formula all the gauge fields associated to the unbroken generators in the following way
\be\label{partial}
A_{\mu}=C_{\mu}M_K + e_{i}^{a}(B_{\mu})^{i}_{j}\bar{e}_{b}^{j} + Z_K D_{\mu}Z_{K}^{\dagger},
\ee
where $C_{\mu}$ is the abelian gauge field associated to $M_K$, whereas $B_{\mu}$ and $D_{\mu}$ are respectively the ${\rm SU}(N+1-K)$ and ${\rm SU}(K)$ connections.
This formula is however incomplete, as it is not form invariant under ``restricted'' gauge transformations. To obtain the full Cho connection we must include the terms
\be
-i\,[\de_{\mu}M_K,M_K] - Z_{\mu},
\ee
where $Z_{\mu}$ is defined as
$$
Z_{\mu}\equiv    i\, Z_K Z_{K}^{\dagger}\de_{\mu}Z_K Z_{K}^{\dagger} +i \, e_{i}^{a} {\bar e}^{i}_{c} \de_{\mu} e_{j}^{c}{\bar e}^{j}_{b}.
$$
The sum of the previous two formulas gives the desired result, as it is form invariant and satisfies the constraint $D_{\mu}M_K=0$, as can be easily checked by direct computation.
Our proposal is therefore
\be\label{CHONA}
A_{\mu}=C_{\mu}M_K + e_{i}^{a}(B_{\mu})^{i}_{j}\bar{e}_{b}^{j} + Z_K D_{\mu}Z_{K}^{\dagger} - Z_{\mu}.
\ee

We will now show that (\ref{CHONA}) is form invariant under restricted gauge transformations. To simplify the formulas we suppress the indices,
hoping that this will not cause misunderstandings.

We recall that $M_K$ is defined as
\be\label{emme}
M_K=Z_K Z_{K}^{\dagger}-\frac{K}{N+1}I_{(N+1)\times(N+1)}= \frac{N+1-K}{N+1}Z_K Z_{K}^{\dagger} - \frac{K}{N+1}e\bar{e}.
\ee
Since the $e_i$ and $u_i$ vectors form an orthonormal basis, we can straightforwardly exponentiate it and deduce that a unitary rotation about the $M_K$
direction with angle $\alpha$ can be written as
$$
U=e\bar{e}\phi^{-\frac{K}{N+1}} + Z_K Z_{K}^{\dagger}\phi^{\frac{N+1-K}{N+1}}, \quad \phi\equiv e^{i\,\alpha}.
$$
We can now easily consider also the ${\rm SU}(N+1-K)$ and ${\rm SU}(K)$ rotations (that we denote with $\Omega_{1}$ and $\Omega_2$ respectively): the general unitary
transformation of the residual gauge group can be parametrized in the following way
\be\label{gauge}
U=e\Omega_{1}\bar{e}\phi^{-\frac{K}{N+1}} + Z_K \Omega_2 Z_{K}^{\dagger}\phi^{\frac{N+1-K}{N+1}}.
\ee

Let us now examine carefully the inhomogeneous term of the gauge transformation (namely $-i\,\de_{\mu}UU^\dagger$):
\be\label{ino}
\de_{\mu}(e\Omega_{1}\bar{e}\phi^{-\frac{K}{N+1}} + Z_K \Omega_2 Z_{K}^{\dagger}\phi^{\frac{N+1-K}{N+1}})(e\Omega_{1}^{\dagger}\bar{e}\phi^{\frac{K}{N+1}} + Z_K \Omega_2^{\dagger} Z_{K}^{\dagger}\phi^{-\frac{N+1-K}{N+1}}).
\ee
We can identify four contributions in the previous formula:
\begin{itemize}
\item When the derivative acts on $\phi$ we find
\be\label{punto1}
\de_{\mu}\alpha\left(\frac{N+1-K}{N+1}Z_K Z_{K}^{\dagger} - \frac{K}{N+1}e\bar{e}\right) = \de_{\mu}\alpha M_K.
\ee
\item When the derivative acts on the $\Omega$'s the result is
\be\label{punto2}
e(-i\,\de_{\mu}\Omega_{1}\Omega_{1}^{\dagger})\bar{e} + Z_K(-i\,\de_{\mu}\Omega_{2}\Omega_{2}^{\dagger})Z_{K}^{\dagger}.
\ee
\item There is another term obtained deriving $e$ and $Z_K$ which has the form
\be\label{punto3}
\de_{\mu}e\bar{e}+\de_{\mu}Z_K Z_{K}^{\dagger}.
\ee
\item The last contribution arises when the derivative acts on $\bar{e}$ and $Z_{K}^{\dagger}$ and it can be rewritten as
\be\label{punto4}
-i\, U(e\de_{\mu}\bar{e} + Z_K \de_{\mu}Z_{K}^{\dagger})U^\dagger.
\ee
\end{itemize}

Notice that $$\de_{\mu}e\bar{e}+\de_{\mu}Z_K Z_{K}^{\dagger} + e\de_{\mu}\bar{e} + Z_K \de_{\mu}Z_{K}^{\dagger} = \de_{\mu}(e\bar{e} + Z_K Z_{K}^{\dagger})= \de_{\mu}I_{(N+1)\times(N+1)} = 0.$$
The first two contributions combine to give the inhomogeneous terms of the gauge transformations for $C_\mu$,
$B_\mu$ and $D_\mu$ whereas the last two terms can be recast in the form
\be\label{trasf}
-i\, (\de_{\mu}e\bar{e}+\de_{\mu}Z_K Z_{K}^{\dagger}) + i\, U(\de_{\mu}e\bar{e}+\de_{\mu}Z_K Z_{K}^{\dagger})U^{\dagger}.
\ee
If we include the term $-i\, (\de_{\mu}e\bar{e}+\de_{\mu}Z_K Z_{K}^{\dagger})$ in the decomposition, the homogeneous part of the gauge transformation will generate new terms that are however cancelled by the corresponding contribution (\ref{trasf}) from the inhomogeneous part; all the other terms just combine to give the gauge transformations for the fields in the unbroken subgroup. We thus only need to prove that the above term gives formula (\ref{CHONA}). If we  multiply it on the left by the matrix $e\bar{e} + Z_K Z_{K}^{\dagger}$, which is the identity, we find
\be
-i\, e\bar{e}\de_{\mu}e\bar{e} -i\, Z_K Z_{K}^{\dagger}\de_{\mu}Z_K Z_{K}^{\dagger}-i\, e\bar{e}\de_{\mu}Z_K Z_{K}^{\dagger} + i\, Z_K \de_\mu Z_{K}^{\dagger}e\bar{e}.
\ee
We can easily see that the last two terms of the previous equation give exactly $-i\,[\de_\mu M_K,M_K]$, thus proving that our connecton is form invariant.

\section{Explicit form of $U$ for embedding without topological obstructions  \label{Noobst}}

The form of the conjugation $U$  needed to give
\be
M=   \frac{1}{|w|^2}  w\overline{w}-\frac{1}{N+1}\mathbf{1},  \qquad  w=\left(\begin{array}{c} 1 \\  v_{1}  \\\vdots \\ v_{N}\end{array}\right) ,
\ee
\be      M =  \sqrt {\frac{2N}{N+1}} \,   U  \, T^{(0)}\,    U^{\dagger},    \qquad   T^{(0)}= \frac{1}{\sqrt{2N(N+1)}}   \left(\begin{array}{cc}N & 0 \\0 & -   {\mathbbm 1}_N\end{array}\right),
\ee
that is,
\be   U   \, \left(\begin{array}{c} 1 \\  0  \\\vdots \\ 0\end{array}\right)   =\frac{1}{|w|} \left(\begin{array}{c} 1 \\  v_{1}  \\\vdots \\ v_{N}\end{array}\right),
\label{impose}\ee
is obtained straightforwardly by exponentiating a matrix of the form
\be      i  ( e A {\bar z} + z A^{\dagger} {\bar e} )^{a}_{b} =   i  ( e^{a}_{j} A^{j} {\bar z}_{b} + z^{a} A^{\dagger}_{j} {\bar e}^{j}_{b} )
\ee
where we use the orthonormal vielbeins introduced in Section~\ref{Chosez}. Actually it is sufficient to use the vielbeins of fixed direction
\be    e^{a}_{j} =\delta^{a}_{j}, \qquad  z^{a}=\delta^{a}_{1}\;. \qquad  j,a=1,2,\ldots, N
\ee
in this calculation.
Calculating
\bqa  U &=&   e^{i  (e A {\bar z} + z A^{\dagger} {\bar e})}  \label{explicitU}   \\
&=&   {\mathbbm 1}_{N+1 \times N+1}  + \left( z{\bar z} + \frac{1}{|A|^{2}}  e A A^{\dagger} {\bar e} \right)
\{ \cos |A| -   1\}  + i   \frac{ \sin |A|}{|A|}    \left( e A {\bar z} + z A^{\dagger} {\bar e}  \right)    \nonumber
\eea
and imposing Eq.~(\ref{impose})  on it  we find
\be  A^{j}  = - i  v_{j} \alpha,
\ee
where   $\alpha>0$ is given  uniquely in terms of $v_{j}$ by
\be        \tan (\alpha  |v|)=  |v|\;.   \label{explicitUBis}
\ee
Eq.~(\ref{explicitU})-Eq.~(\ref{explicitUBis}) give the desired expression of $U$.

\begin{thebibliography}{23}


\bibitem{FN1}
  L.~D.~Faddeev and A.~J.~Niemi,
  ``Partially dual variables in SU(2) Yang-Mills theory,''
  Phys.\ Rev.\ Lett.\  {\bf 82} (1999) 1624
  [arXiv:hep-th/9807069].

\bibitem{FN2}
  L.~D.~Faddeev and A.~J.~Niemi,
  ``Partial duality in SU(N) Yang-Mills theory,''
  Phys.\ Lett.\  B {\bf 449} (1999) 214
  [arXiv:hep-th/9812090].

\bibitem{thooft}
  G.~'t Hooft,
  ``Topology Of The Gauge Condition And New Confinement Phases In Nonabelian
  Gauge Theories,''
  Nucl.\ Phys.\  B {\bf 190} (1981) 455.

\bibitem{nodi1}
  L.~D.~Faddeev and A.~J.~Niemi,
  ``Knots and particles,''
  Nature {\bf 387} (1997) 58
  [arXiv:hep-th/9610193].

\bibitem{nodi2}
  L.~D.~Faddeev and A.~J.~Niemi,
  ``Toroidal configurations as stable solitons,''
  [arXiv:hep-th/9705176].


\bibitem{noi}
  J.~Evslin and S.~Giacomelli,
  ``A Faddeev-Niemi Solution that Does Not Satisfy Gauss' Law,''
  arXiv:1010.1702 [hep-th].

\bibitem{Niemi}
  A.~J.~Niemi and A.~Wereszczynski,
  ``On Solutions to the 'Faddeev-Niemi' Equations,''
  arXiv:1011.6667 [hep-th].

\bibitem{Shabanov}
 S.~V.~Shabanov,
  ``An effective action for monopoles and knot solitons in Yang-Mills
  theory,''
  [arXiv:hep-th/9903223];
 S.~V.~Shabanov,
  ``Yang-Mills theory as an Abelian theory without gauge fixing,''
  [arXiv:hep-th/9907182].

\bibitem{BCK}
  W.~S.~Bae, Y.~M.~Cho and S.~W.~Kim,
  ``QCD versus Skyrme-Faddeev theory,''
  Phys.\ Rev.\  D {\bf 65} (2002) 025005
  [arXiv:hep-th/0105163].

\bibitem{Kondo}
  K.~I.~Kondo, T.~Murakami and T.~Shinohara,
  ``Yang-Mills theory constructed from Cho-Faddeev-Niemi decomposition,''
  [arXiv:hep-th/0504107];
   K.~I.~Kondo, T.~Murakami and T.~Shinohara,
  ``BRST symmetry of SU(2) Yang-Mills theory in Cho-Faddeev-Niemi
  decomposition,''
  [arXiv:hep-th/0504198].

\bibitem{FNvinc}
  L.~D.~Faddeev and A.~J.~Niemi,
  ``Decomposing the Yang-Mills field,''
  [arXiv:hep-th/9907180].


\bibitem{CDyons} A. ~Abouelsaood,  Nucl. Phys.  B {\bf 226},  309 (1983);  P. ~Nelson, A. ~Manohar,  Phys. Rev. Lett. {\bf  50},  943
(1983);  A. ~Balachandran, G. ~Marmo, M. ~Mukunda, J. ~Nilsson, E. ~Sudarshan, F. ~Zaccaria,    Phys. Rev. Lett. {\bf  50},  1553
(1983);  P. ~Nelson, S. ~Coleman,  Nucl. Phys.  B {\bf 227},  1  (1984)

\bibitem{DFHK} N. ~Dorey, C. ~Fraser, T.J. ~Hollowood,  M.A.C.~ Kneipp,
 ``NonAbelian duality in N=4 supersymmetric gauge theories,''  Phys.Lett.  B {\bf 383}, 422  (1996)  [arXiv: hep-th/9512116].

  \bibitem{susy}
P. C. ~Argyres, M. R. ~Plesser,  N. ~Seiberg,
``The Moduli space of vacua of N=2 SUSY QCD and duality in N=1 SUSY QCD'',
 Nucl. Phys.  B {\bf 471}, 159
(1996) [arXiv: hep-th/9603042];   K.  ~Hori, H. ~Ooguri,   Y.  ~Oz,
``Strong coupling dynamics of four-dimensional N=1 gauge theories from M theory five-brane,''
 Adv. Theor. Math. Phys.   {\bf 1}, 1  (1998)  [arXiv:  hep-th/9706082];
 A. ~Hanany,  Y. ~Oz,
 ``On the quantum moduli space of vacua of N=2 supersymmetric SU(N(c)) gauge theories,''
  Nucl. Phys.  B {\bf  452},  283  (1995) [arXiv: hep-th/9505075];
G. ~Carlino, K. ~Konishi,  H. ~Murayama,
``Dynamical symmetry breaking in supersymmetric SU(n(c)) and USp(2n(c)) gauge theories,''
 Nucl. Phys.   B {\bf 590},  37   (2000) [arXiv: hep-th/0005076];
G. ~Carlino, K. ~ Konishi, S. P. ~ Kumar, H. ~ Murayama,
``Vacuum structure and flavor symmetry breaking in supersymmetric SO(n(c)) gauge theories,''
Nucl.  Phys.  B {\bf 608}, 51 (2001)  [arXiv: hep-th/0104064].

\bibitem{noimono}
    R.~Auzzi, S.~Bolognesi, J.~Evslin and K.~Konishi,
  ``NonAbelian monopoles and the vortices that confine them,''
  Nucl.\ Phys.\  {\bf B686 } (2004)  119   [arXiv: hep-th/0312233];

\bibitem{Duality} M. ~Eto et. al.,
``Non-Abelian duality from vortex moduli: A Dual model of color-confinement,''
Nucl. Phys.   B {\bf 780},  161  (2007) [arXiv: hep-th/0611313];
 K.~Konishi,
 ``The Magnetic Monopoles Seventy-Five Years Later'',
Lecture Notes in Physics, 737: 471-521, (2008),
[arXiv: hep-th/0702102].


\bibitem{Confind} K. ~Konishi and Y.~Ookouchi,
``On Confinement Index,''
Nucl. Phys.   B {\bf 827}, 59  (2010),  arXiv:0909.3781 [hep-th].

\bibitem{DouglasShenker}
  M.~R.~Douglas and S.~H.~Shenker,
  ``Dynamics of SU(N) supersymmetric gauge theory,''
  Nucl.\ Phys.\  B {\bf 447}, 271 (1995)
  [arXiv:hep-th/9503163].

\bibitem{HananyZaffaroni}
  A.~Hanany and A.~Zaffaroni,
  ``On the realization of chiral four-dimensional gauge theories using branes,''
  JHEP {\bf 9805} (1998) 001
  [arXiv:hep-th/9801134].

\bibitem{ABEKY}
  R.~Auzzi, S.~Bolognesi, J.~Evslin, K.~Konishi and A.~Yung,
  ``Nonabelian superconductors: Vortices and confinement in N = 2 SQCD,''
  Nucl.\ Phys.\  B {\bf 673} (2003) 187
  [arXiv:hep-th/0307287].

\bibitem{FNZwei}
L.D.~Faddeev and A.J.~ Niemi, 
``Aspects of electric magnetic duality in SU(2) Yang-Mills theory'', 
Phys. \ Lett. \ B {\bf 525} (2002) 195 
[arXiv: hep-th/0101078];
T.A.~Bolokhov and L.D.~Faddeev, 
``Infrared variables for the SU(3) Yang-Mills fields'',
Theor. and Mathem. Physics, 139 (2004), 679 
(translated from Theoreticheskaya i Matematicheskaya Fizika, 139 (2004), 276.)

\bibitem{DuanGe}
Y.~S.~Duan and M.~L.~Ge,
Sci. Sin. {\bf 11} (1979) 1072.

\bibitem{Cho1}
  Y.~M.~Cho,
  ``A Restricted Gauge Theory,''
  Phys.\ Rev.\  D {\bf 21} (1980) 1080.

\bibitem{Cho2}
  Y.~M.~Cho,
  ``Extended Gauge Theory And Its Mass Spectrum,''
  Phys.\ Rev.\  D {\bf 23} (1981) 2415.


\bibitem{Kondonew}
K. ~Kondo, A.~Shibata, T.~Shinohara and S.~Kato,
 ``Non-Abelian magnetic monopoles responsible for quark confinement,''
 	arXiv:1102.4150v1 [hep-th].



\bibitem{Wuyang}  T.T. Wu,  C.N. Yang,  in  ``Properties of Matter Under
Unusual Conditions", Ed. H. Mark, S. Fernbach, Interscience, New York (1969).

\bibitem{KonTak}   K. Konishi, K. Takenaga,  Phys. Lett. B {\bf 508}, 392 (2001) [hep-th/9911097].

\bibitem{Sigma}
  F.~Delduc and G.~Valent,
  ``Renormalizability Of The Generalized Sigma Models Defined On Compact Hermitian Symmetric Spaces,''
  Nucl.\ Phys.\  B {\bf 253} (1985) 494.

\end{thebibliography}
\end{document}